\documentclass[journal]{IEEEtran}

\IEEEoverridecommandlockouts %/thanks
\usepackage{graphicx,latexsym,subfigure,amsmath,epsfig,latexsym,subfigure,amsmath,cite,amssymb}
\usepackage{amsthm,mathrsfs}
\usepackage{graphicx}
\usepackage{epstopdf}
\usepackage{color}
\usepackage{CJK}
\usepackage[ruled,linesnumbered]{algorithm2e}
\usepackage{mathrsfs}%花体设置
\usepackage{bm}%希腊字母加黑
\usepackage{float}%调整图片位置
\usepackage{setspace}%使用间距宏包
\usepackage{color}%使用颜色宏包
\usepackage{subfigure}%插入并排图片
\usepackage{balance}%均衡最后一页内容
\usepackage{subeqnarray}%大括号公式编号
\usepackage{cases}
\begin{document}
\title{Security in Mobile Edge Caching with Reinforcement Learning}
\author{\IEEEauthorblockN{\footnotesize Liang Xiao\IEEEauthorrefmark{1}, Xiaoyue Wan\IEEEauthorrefmark{1}, Canhuang Dai\IEEEauthorrefmark{1},
Xiaojiang Du\IEEEauthorrefmark{2},
Xiang Chen\IEEEauthorrefmark{3},
Mohsen Guizani\IEEEauthorrefmark{4}}\\
\IEEEauthorblockA{\IEEEauthorrefmark{1}\footnotesize Dept. of Communication Engineering, Xiamen Univ., Xiamen, China. Email: lxiao@xmu.edu.cn}\\
\IEEEauthorblockA{\IEEEauthorrefmark{2}Dept. of Computer and Information Science, Temple Univ., Philadelphia, USA. Email: dxj@ieee.org}\\
\IEEEauthorblockA{\IEEEauthorrefmark{3}Dept. of Electronics and Information Tech., Sun Yat-sen Univ., Guangzhou, Guangdong Province, China. Email: chenxiang@mail.sysu.edu.cn}\\
\IEEEauthorblockA{\IEEEauthorrefmark{4}Dept. of Electrical and Computer Engineering, Univ. of Idaho, Moscow, Idaho, USA. Email: mguizani@ieee.org}
}

\maketitle
\begin{abstract}
%Mobile edge computing (MEC) systems with caching that support multimedia contents in 5G mobile Internet with low latency and computing overhead have to address various attacks such as denial of service attacks and rogue edge attacks. This article investigates the attack models in mobile edge computing systems, focusing on both the mobile offloading and the caching procedures. The security solutions are proposed, in which reinforcement learning (RL) techniques are applied to provide secure offloading to the edge nodes against smart attacks, light-weight authentication and secure collaborative caching to protect data privacy. We evaluate the performance of the RL-based security solution for mobile edge caching and discuss the challenges that have to the addressed in the future.
Mobile edge computing usually uses cache to support multimedia contents in 5G mobile Internet to reduce the computing overhead and latency. Mobile edge caching (MEC) systems are vulnerable to various attacks such as denial of service attacks and rogue edge attacks. This article investigates the attack models in MEC systems, focusing on both the mobile offloading and the caching procedures. In this paper, we propose security solutions that apply reinforcement learning (RL) techniques to provide secure offloading to the edge nodes against jamming attacks. We also present light-weight authentication and secure collaborative caching schemes to protect data privacy. We evaluate the performance of the RL-based security solution for mobile edge caching and discuss the challenges that need to be addressed in the future.
\end{abstract}


\IEEEpeerreviewmaketitle
\section{Introduction}
\begin{figure*}[!htbp]
\centering\includegraphics[width=6.5in,height=2.7in]{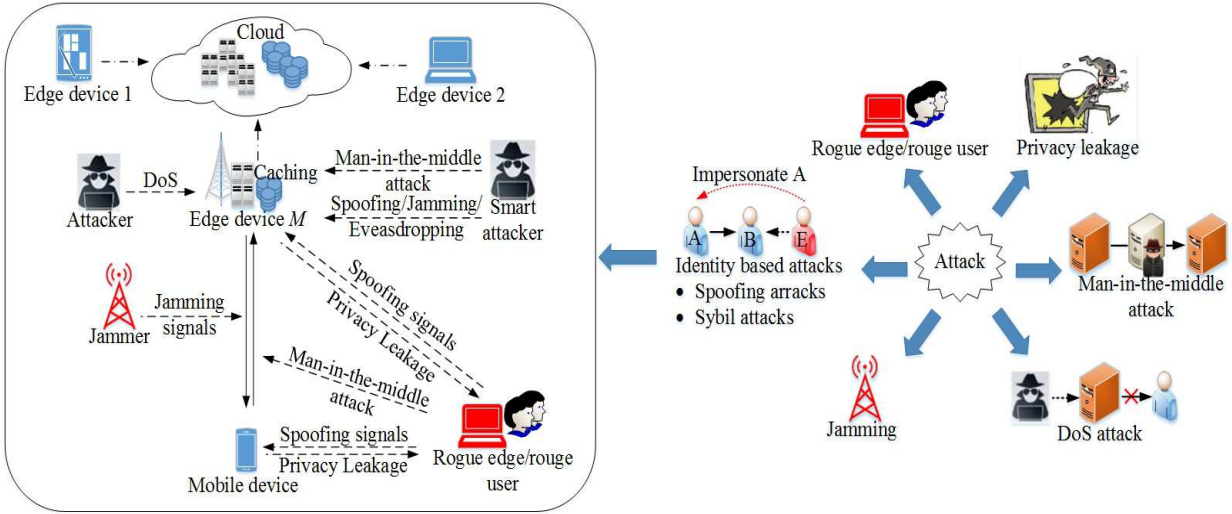}\\
\caption{Threats in mobile edge caching.}\label{fig:threat}
\end{figure*}
Mobile edge computing provides data storage, computing and application services with edge devices such as access points (APs), laptops, base stations, switches and IP video cameras at the network edge. Being closer to customers than cloud, mobile edge computing can support the Internet of Things (IoT), cyber-physical systems, vehicular networks, smart grids and embedded AI with lower latency, location awareness and mobility support \cite{Mao2017A}. {\color{black} Mobile edge caching reduces the duplicated transmissions and backhaul traffic, improves the communication efficiency, and provides quality of services for caching users.} Collaborative caching in mobile edge computing shares popular data such as multimedia contents in video games with augmented reality among end users and significantly reduces the traffic load and service latency in the 5G mobile Internet \cite{Sun2016EdgeIoT}.

Security and data privacy are critical and become the bottleneck for the development of mobile edge caching (MEC), as edge devices are located at the edge of the heterogenous networks and physically closer to attackers. With limited computation, energy, communication and memory resources, the edge devices are protected by different types of security protocols, which are in general less secure compared with cloud servers and data centers. In addition, mobile edge caching systems consist of distributed edge devices that are controlled by selfish and autonomous people. The edge device owners might be curious about the data contents stored on their cache and sometimes even launch insider attacks to analyze and sell the privacy information of the customers. Therefore, MEC systems are more vulnerable to security threats such as wireless jamming, distributed denial of service attacks (DoS), spoofing attacks including rogue edge and rogue mobile devices, man-in-the-middle attacks, and smart attacks \cite{Montero2015Virtualized,chiang2016fog,Roman2016Mobile}.

In this article, we briefly review the security and privacy challenging of mobile edge caching and investigate the tradeoff between the MEC security performance and the protection overhead in terms of the computation complexity and time, communication overhead and energy consumption. Edge devices and mobile devices have different computing and storage resources, battery levels, communication bandwidths and locations.  Each node has to optimize its defense strategy and choose the key parameters in the security protocols, which are challenging in the heterogenous dynamic network as the dynamic network model and attack model are difficult to estimate. For instance, the test threshold as a key parameter in the PHY-authentication is set based on the known radio propagation and spoofing model, or a large number of training data. However, neither the network model or the large volume of training data can be readily obtained in time for an edge node or mobile device to authenticate each received message \cite{Xiao2016PHY}.

The dilemma in MEC security can be addressed by reinforcement learning (RL) techniques, which enable a learning agent to derive an ``optimal" strategy via trial-and-error. It has been proved that Q-learning, the model-free and widely-used RL algorithm can achieve the highest cumulative reward in the Markov decision process (MDP) \cite{jiang2017machine}. By applying RL techniques, cyber systems such as AlphaGo have beaten human players in various games and have attracted extensive attentions from both academia and industry. In recent years, RL techniques have been used to study the dynamic security games, and the proposed RL-based security schemes such as the anti-jamming channel access scheme, the authentication scheme and the malware detection scheme exceed the benchmark deterministic schemes \cite{Aref2017Jamming,Li2016malware,Xiao2016PHY,Xie2016User,wu2012anti}. Therefore, we investigate the repeated game between the MEC systems and attackers and discuss how to build the RL-based security solutions, such as the secure mobile offloading against jamming and smart attacks, the light-weight authentication with multiple protection levels and collaborative caching to resist eavesdropping.

We briefly review the RL-based security techniques and compare their performance via simulations. The challenges to implement the RL-based edge security solutions on practical mobile edge caching are discussed. Developed mostly for games such as Go and video games, most reinforcement learning techniques require the agent to accurately observe the environment state and receive an immediate reward from each action. Unfortunately, these conditions rarely hold in the MEC security game and MEC systems have to be protected from the security disasters due to the trial errors of the RL algorithms.

This article is organized as follows. In the next section we review the main security issues and present the attack models in MEC systems. We then describe how to build the RL-based MEC security solutions and evaluate their performance. We also identify the challenges ahead and point out several possible directions for future work. Finally, we draw conclusions.

\section{Threat Model in Mobile Edge Caching}

\begin{figure*}[!htbp]
\centering\includegraphics[width=5.0in,height=3.2in]{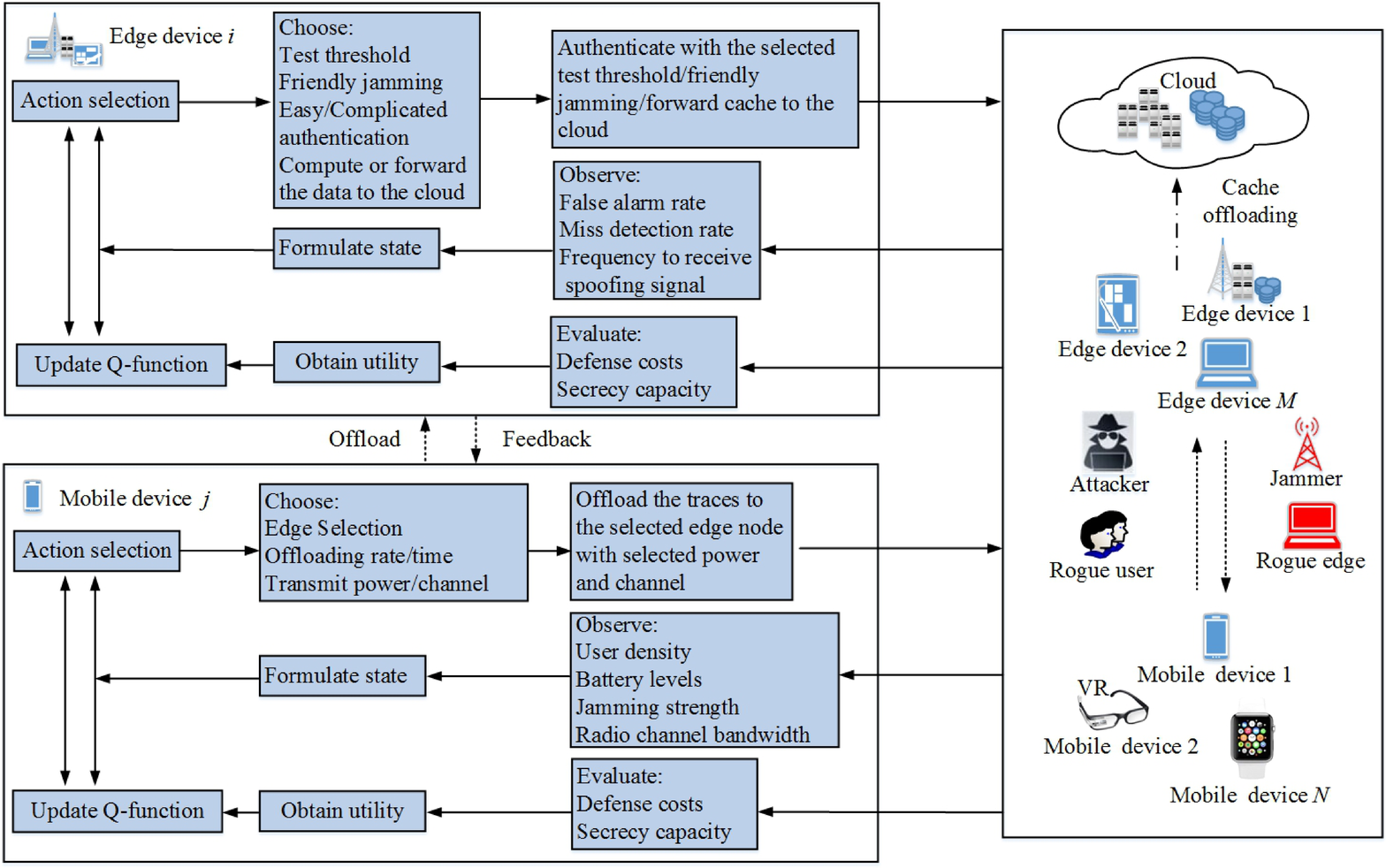}\\
\caption{RL-based security solutions for MEC systems.}\label{fig:secure_offloading}
\end{figure*}
%\begin{figure*}[!ttbp]
%\centering
%\subfigure[Mobile device]{
%\label{fig:RL_device}
%\includegraphics[height=2.1 in]{F2a.eps}}\\
%%\hspace{0.5in}
%\subfigure[Edge device]{
%\label{fig:RL_edge}
%\includegraphics[height=2.1 in]{F2b.eps}}\\
%\caption{RL-based security solutions for MEC systems.}
%\label{fig:secure_offloading}
%\end{figure*}

In mobile edge caching, adversary can compromise a number of ``weak" edge nodes such as video cameras that are only protected with light-weight authentication and encryptions. By using the compromised edge nodes and/or commercial radio devices such as laptops, an attacker can attack the mobile devices and/or edge nodes. In addition, selfish customers and curious owners of the edge devices who are hunger for secrets and money also have motivations to attack MEC systems, if they know their illegal gains do not incur any punishment. Moreover, by applying advanced machine learning techniques and smart radio transmission methods, a smart attacker can learn the ongoing network status and chooses its attack strategy accordingly and flexibly in each time slot, which makes it more dangerous for MEC systems.

As illustrated in Fig. \ref{fig:threat}, mobile edge caching has to address a large number of attacks during the mobile offloading procedure and the caching perspective. During mobile offloading, the radio communication channels of an MEC system are vulnerable to the attacks launched from the physical layer or MAC layers, such as jamming, rogue edge nodes/mobile devices, eavesdropping, man-in-the-middle attacks and smart attacks. The data stored in the cache of the edge devices have to be protected to avoid privacy leakage. We briefly review some important types of attacks as follows.

\textbf{Jamming:} {\color{black} A jammer sends faked signals to interrupt the ongoing radio transmissions of the edge node with cached chunks or caching users and prevent the caching users to access the cached contents. Another goal of the jammers is to deplete the bandwidth, energy, central processing unit (CPU) and memory resources of the victim edge nodes, caching users and sensors during their failed communication attempts \cite{Han2017Two}.}

\textbf{DoS:} DoS attacks are one of the most dangerous security threats, in which attackers aim to break down the victim computer network or cyber systems and interrupt their services. MEC systems are especially vulnerable to distributed DoS attacks, in which some distributed edge devices that are not well protected by security protocols can be easily compromised and then used to attack other edge nodes. {\color{black} Some attackers also aim to prevent the collaborative caching users from accessing the caching data.} Jamming can be viewed as a special type of DoS attacks.

\textbf{Spoofing attacks/Rogue edge/Rogue mobile user/Sybil attacks:} {\color{black} An attacker sends spoofing signals to edge nodes with cached chunks or the caching users with the identity of another node such as the MAC address to obtain illegal access of the network resources, and perform further attacks such as DoS and man-in-the-middle attacks \cite{Xiao2016PHY}. For example, an attacker claims to be an edge node to fool the mobile devices in the area in rogue edge attacks, or sends spoofing messages to the edge node with the identity of another user in rogue user attacks. Faked caching space claimed by the rogue edge can result in significant data loss among the caching users in the collaborative MEC shared with a large number of users. In Sybil attacks as another type of identity-based attacks, a caching user claims to be multiple users and request more network and storage resources.}

\textbf{Man-in-the-middle attacks:} {\color{black} Man-in-the-middle attacker sends jamming and spoofing signals to fake an edge node \cite{Roman2016Mobile} with the goal of hijacking the private communication of the victim edge nodes or mobile devices and even control them.}

\textbf{Privacy leakage:} {\color{black} Some owners of the edge devices are curious about the data stored in their caching and apply machine learning techniques and data analysis software to scan the caching data. In addition, the light-weight authentication protocols cannot always prevent rogue caching users from accessing the caching data. Therefore, MEC systems have to protect the caching user privacy information such as the preferences and travel histories of a specific user during the mobile offloading and the caching process.}

\textbf{Smart Attacks:}
By using smart radio devices such as universal software radio peripherals (USRPs), an attacker can observe the network state such as the traffic pattern in the area, compromise some edge nodes with insufficient security protections, and wiretap the public control channels of the edge network. The attacker can also use machine learning techniques to investigate the network pattern and attack the MEC systems accordingly and possibly with multiple steps. For example, a proactive eavesdropper may first send jamming or spoofing signals to the victim edge node to receive more information from it. In \cite{Xiao2016A}, a smart attacker can choose the type of the attacks according to its distances to the edge nodes, which has been proved to be more dangerous to MEC systems than the traditional attackers that can launch a single type of attacks.

\section{RL-based MEC Security Solutions}
%%%%%%%%%%%%%%%%%%%%%%%%%%%%%%%%%%%%%%%%%%
\begin{table*}[!htbp]
  \caption{Summary of the RL-based security methods in wireless networks}
\newcommand{\tabincell}[2]{\begin{tabular}{@{}#1@{}}#2\end{tabular}}
  \centering
  \begin{tabular}{|c|c|c|c|c|}\hline
Attack & RL techniques & Action & Performance & Ref\\\hline
Spoofing & \tabincell{l}{Q-learning\\Dyna-Q\\DQN} &\tabincell{l}{Test threshold\\Offloading rate\\Auth. level}& \tabincell{l}{False alarm rate\\Miss detection rate\\Utility of the receiver} & \cite{Xiao2016PHY,Xiao2016A}\\\hline
\tabincell{l}{Jamming} & \tabincell{l}{Q-learning\\PDS\\Hotbooting Q\\DQN\\Fast DQN} & \tabincell{l}{Channel selection\\Power control\\Offloading rate}& \tabincell{l}{SINR\\BER\\Energy consumption}& \cite{wu2012anti,Aref2017Jamming,Xiao2016A,Xie2016User,Han2017Two,gwon2013competing}\\\hline
Eavesdropping & \tabincell{l}{Q-learning\\DQN\\Fast DQN} & \tabincell{l}{Defense mode\\Offloading rate}& Secrecy data rate & \cite{Xie2016User}\\\hline
Malware& \tabincell{l}{Q-learning\\Dyna-Q\\PDS} & Offloading rate & \tabincell{l}{Detection accuracy\\Detection delay} & \cite{Li2016malware}\\\hline
\end{tabular}\label{tabb}
\end{table*}
Each edge device or mobile device in MEC systems has to make a number of decisions to address the security threats mentioned in the previous section. For instance, a mobile device has to choose the data, the transmit power, channel and time, and the edge node in the mobile offloading against smart attackers who launch jamming, eavesdropping, rogue edge, and man-in-the-middle attacks according to the ongoing offloading policies and the network states. Most existing edge security solutions are either fixed strategies based on a certain fixed network and attack model or the optimization results based on the accurate knowledge on a number of parameters that are challenging to be obtained by an edge node in a practical edge system, because many of the network and attack parameters change significantly over time and are difficult to be estimated. Therefore, an MEC system has to
find a proper security strategy without heavily depending on a specific network and attack model, which cannot be formulated as an optimization problem that is easy to address by an edge node or mobile device.

This dilemma is promising to be addressed by applying reinforcement learning techniques such as deep Q-network (DQN) and RL-based security solutions enable a wireless device to optimize its policy in the repeated security game via trial-and-error. In the RL-based security scheme, a learning agent such as an edge node or mobile device observes the current state and a quality function or Q-function to choose its action such as the security complexity and defense levels. The state corresponds to the status of the other nodes in the MEC system and the attack characters that can be observed by the node. If the future reward to the node is independent of the previous state for the given current state and strategy, the node can achieve the optimal strategy after sufficient interactions with the attackers in the dynamic edge system.

One of the first wireless security issues that apply reinforcement learning techniques is anti-jamming communications \cite{wu2012anti,Aref2017Jamming,Xiao2016A,Xie2016User,Han2017Two,gwon2013competing}, showing that a transmitter can use RL algorithms such as Q-learning to optimize its transmit power and channel selection in some simplified communication scenarios, such as very few number of feasible actions and possible states, without being aware of the network model and the jamming model. As summarized in Table \ref{tabb}, the RL techniques have also been used in spoofing detection\cite{Xiao2016PHY,Xiao2016A}, smart attacks \cite{Xiao2016A} and malware detection \cite{Li2016malware}. Therefore, reinforcement learning is promising to improve MEC security, although the RL-based MEC security solutions are complicated with more challenges to address. For concrete examples, we show how to apply RL techniques in the anti-jamming offloading, authentication and anti-eavesdropping transmission issues as follows.

\textbf{RL-based anti-jamming mobile offloading:} In an MEC system, a mobile device has to choose its offloading policy, such as the part of the data to offload, the transmit power, channel and time, and which edge nodes to connect to, each from a given finite feasible action set. The goal is to improve the offloading quality such as the signal-to-noise-plus-interference (SINR) and bit error rate (BER) of the signals received by the edge nodes against jamming and interference and save the computation and communication energy consumption.

{\color{black} As the future state observed by a mobile device is independent of the previous states and actions for a given state and offloading strategy in the current time slot, the mobile offloading strategy chosen by the mobile device in the repeated game with jammers and interference sources can be viewed as a MDP with finite states \cite{Xiao2016A}. Therefore, a mobile device can apply reinforcement learning techniques to achieve the optimal offloading policy without being aware of the jamming model and the MEC model.}
%The mobile offloading in the repeated game between the mobile device and the network including the edge nodes, the jammers and interference can be viewed as a MDP with finite states.

{\color{black} In the RL-based offloading scheme as presented in \cite{Li2016malware}, the mobile device observes received jamming power, the radio channel bandwidth, the battery levels and the user density to formulate the state. As illustrated in Fig. \ref{fig:secure_offloading}, the mobile device chooses the offloading policy such as the edge selection and offloading rate based on the current state and the Q-function, which is the expected discounted long-term reward for each action-state pair and represents the knowledge obtained from the jamming defense history.
%The Q-learning based offloading scheme uses the iterative Bellman equation to update the Q-function.
The values of the Q-function are updated via the iterative Bellman equation in each time slot according to the current offloading policy, the network state and the utility received by the mobile device against jamming.}

\begin{figure*}[!t]
\centering\includegraphics[width=4.7in,height=3.0in]{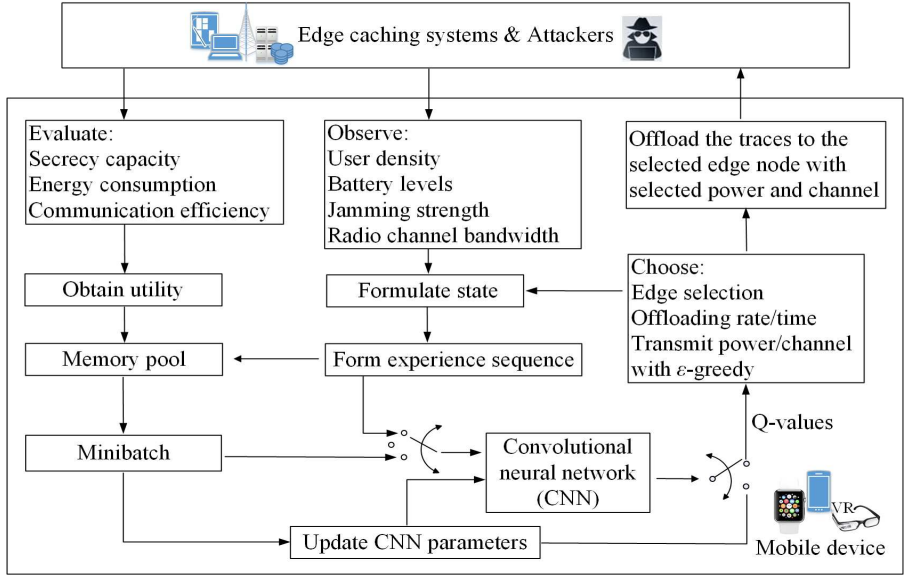}\\
\caption{Illustration of the DQN-based secure offloading in mobile edge caching.}\label{fig:DQN_secure_offloading}
\end{figure*}

{\color{black} The utility of the mobile device received in a time slot is evaluated according to the anti-jamming communication efficiency such as the SINR of the signals, the BER of the received messages and the defense costs such as the offloading energy consumption.
%The Q-learning based offloading scheme uses the iterative Bellman equation to update the Q-function.
In the DQN based offloading scheme, convolutional neural networks (CNN) and the strategy sequence pool as shown in Fig. \ref{fig:DQN_secure_offloading} are used to estimate the Q-values and provide a faster learning speed.
The CNN consists of two convolutional (Conv) layers and two fully connected (FC) layers, and the weights of the CNN are updated based on the stochastic gradient descent (SGD) algorithm according to the previous anti-jamming communication experience in the memory pool \cite{Han2017Two}. The output of the CNN is used for estimating the values of the Q-function for each offloading policy.
By applying the $\epsilon$-greedy algorithm, the mobile device chooses the offloading policy that maximizes its current Q-function with a high probability $1-\epsilon$ and the other policies with a small probability. This scheme can make a tradeoff between the exploration (i.e., to avoid being trapped in the local optimal strategy) and the exploitation (i.e., to improve the utility).}
%By applying the $\epsilon$-greedy algorithm, the mobile device tends to choose the offloading policy that maximizes its Q-function, which makes a tradeoff between exploration (i.e., to avoid being trapped in the local optimal strategy) and exploitation (i.e., to improve the utility).

\textbf{RL-based authentication:} Due to the limited memory, energy and computational resources, a mobile device usually has difficulty estimating the ongoing spoofing model and prefers the light-weight authentication protocols to detect the identity-based attacks such as spoofing attacks, Sybil attacks and rogue edge attacks. Each edge node also needs the fast detection of a large number of spoofing messages and rogue users. To this end, PHY-authentication techniques that reuse the existing channel estimates of the source node and/or the ambient radio signals provide light-weight protection against identity-based attacks without leaking user privacy such as their locations \cite{Xiao2016PHY}.

However, most existing PHY-authentication builds hypothesis tests to compare the radio channel with the channel record of the claimed node. Therefore, the receiver has to determine the test threshold in the authentication for each incoming message, which is challenging in the mobile edge caching system with time-variant radio channel model and spoofing model. This issue can be addressed by RL-based authentication schemes, in which the key authentication parameters such as the test threshold are obtained via reinforcement learning techniques. For example, according to the RL-based authentication scheme as developed in \cite{Xiao2016PHY}, an edge node observes the recent spoofing detection accuracy and the spoofing frequency and chooses the test threshold according to the Q-function which is updated similar to the anti-jamming offloading mentioned above. In another example, similar to \cite{Xiao2016A}, RL techniques can be used for an edge node to determine its authentication methods, i.e., the edge node automatically applies more authentication protocols if finding itself in a risky network with smart attackers.

%%%%%%%%%%%%%%%%%%%%%%%%%%%%%%%%%%%%
%%%%%%%%%%%%%%%%%%%%%%%%%%%%%%%%%%%%%
%\begin{figure*}[!t]
%\centering\includegraphics[width=5.7in,height=1.7in]{authentication_collaborative_caching_v0827_1.eps}\\
%\caption{Illustration of authentication and secure collaborative caching with RL in mobile edge computing.}\label{fig:authentication_collaborative_caching}
%\end{figure*}

\textbf{RL-based friendly jamming:} Secure collaborative caching in MEC has to protect data privacy and resist eavesdropping. For example, an edge node can send friendly jamming signals according to the data stored in the caching system to prevent the eavesdropping attacker from understanding the information sent from a mobile node or another edge node. In this way, each edge node has to determine whether to attend the friendly jamming according to the network topology, the channel models and the presence of the attackers. An edge node has to decide whether to compute the data or to forward the data received from the mobile device to the cloud, and whether to store the ``popular" data in the edge against privacy leakage and DoS attacks.

\section{Performance Analysis}
%\begin{figure*}[!ttbp]
%\centering
%\subfigure[Local energy consumption]{
%\label{fig:energy}
%\includegraphics[height=2.1 in]{F4a.eps}}
%%\hspace{0.5in}
%\subfigure[Computing delay]{
%\label{fig:delay}
%\includegraphics[height=2.1 in]{F4b.eps}}
%\caption{Performance of the RL-based offloading for a mobile device that is close to 3 edge devices against jamming.}
%\label{fig:ave_performance}
%\end{figure*}
\begin{figure*}[!ttbp]
\centering
\subfigure[SINR]{
\label{fig:SINR}
\includegraphics[height=1.7 in]{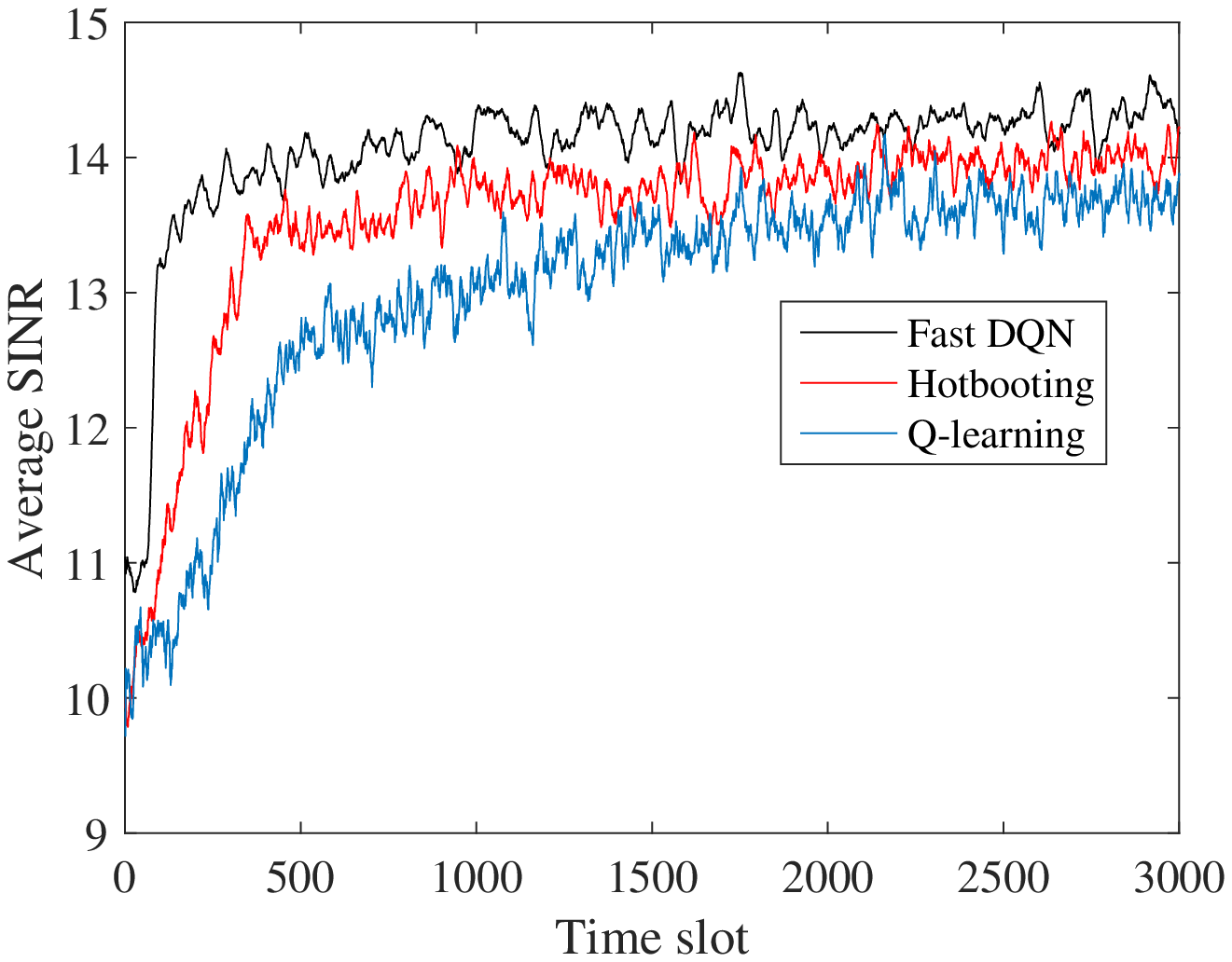}}
\subfigure[Local energy consumpation]{
\label{fig:energy}
\includegraphics[height=1.7 in]{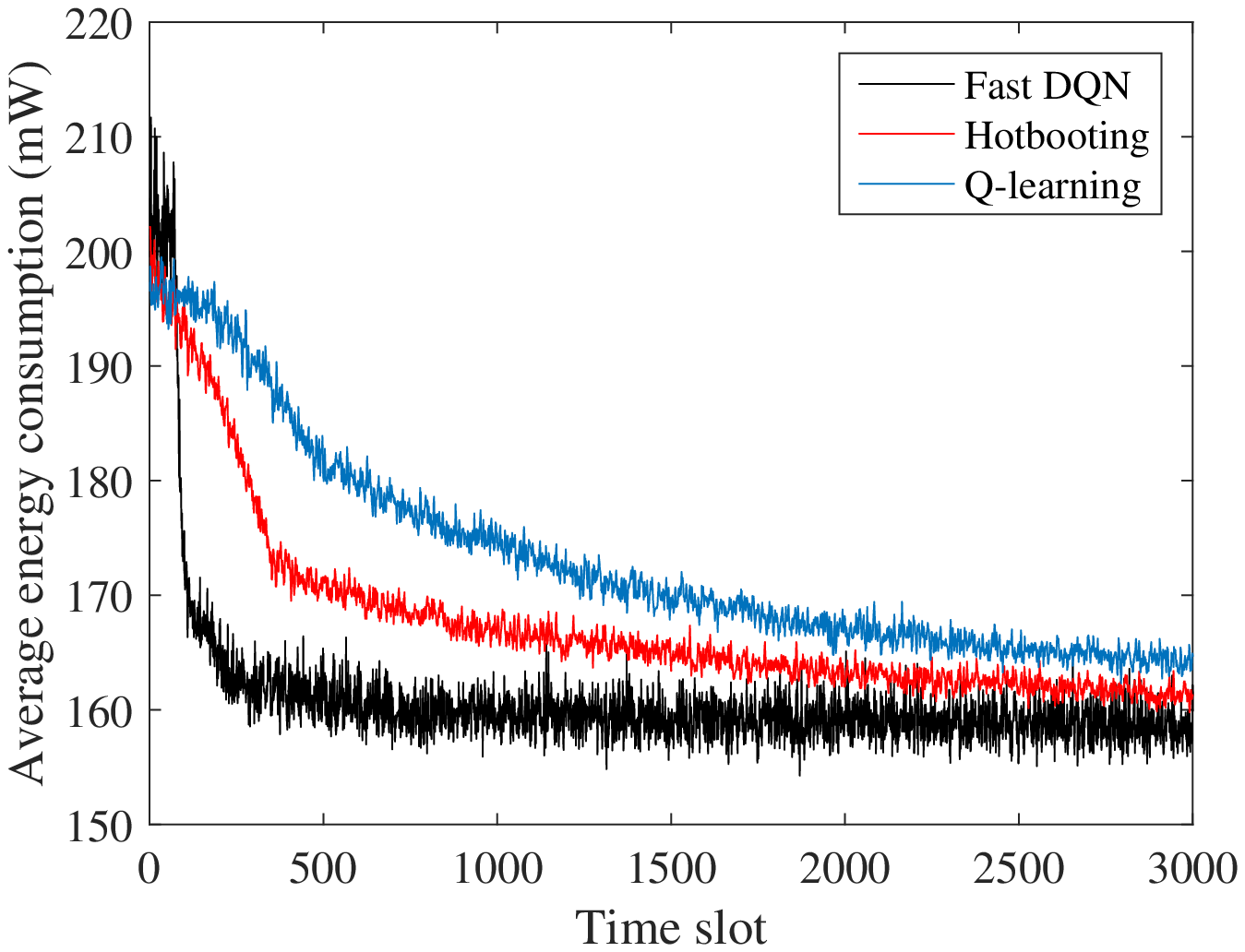}}
%\hspace{0.5in}
\subfigure[Computing delay]{
\label{fig:delay}
\includegraphics[height=1.7 in]{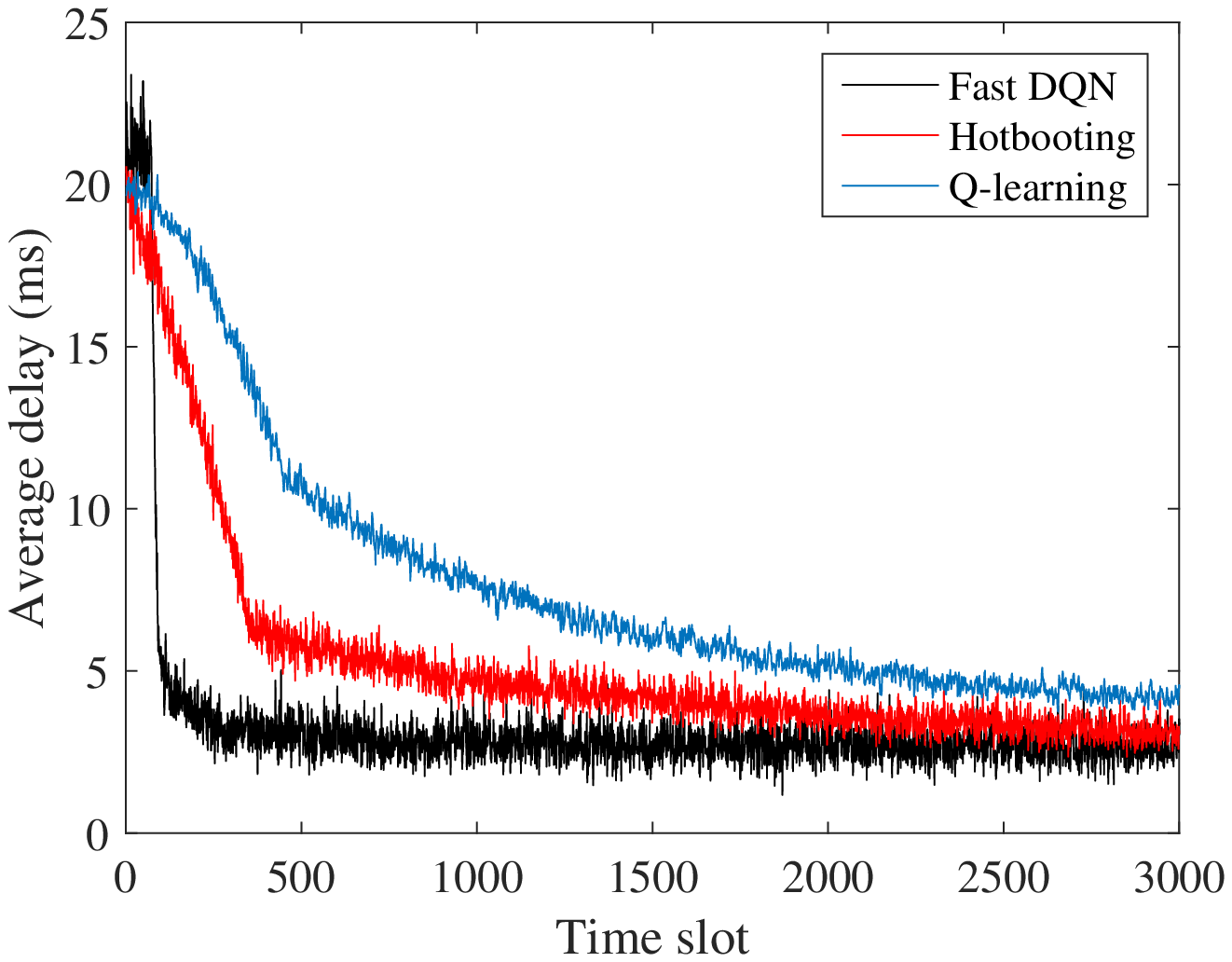}}
\caption{Performance of the RL-based offloading for a mobile device that is close to 3 edge devices against jamming.}
\label{fig:ave_performance}
\end{figure*}
As a widely used RL algorithm, Q-learning has been applied to resist spoofing, malware, jamming and eavesdropping in wireless networks as summarized in Table \ref{tabb}. Without requiring any knowledge on the network and attack model, the Q-learning based security schemes apply the iterative Bellman equation to update the Q-values, and have two parameters, i.e., the learning rate and discount factor to control their learning performance. {\color{black} More specifically, the learning rate is set to weight the current experience in the learning process and the discount factor represents the uncertainty on the feature rewards. In the Q-learning based authentication scheme presented in \cite{Li2016malware}, the learning rate is set as 0.7 and the discount factor is 0.1 to achieve accurate spoofing detection.} These schemes can be easily implemented in mobile or edge node with low computational and storage overhead, and enable it to achieve the optimal strategy with probability one after a sufficiently large number of interactions with the attackers in a MDP even with randomness.

However, the Q-learning based edge security suffers from the ``high-dimensional disaster", as the mobile or edge node has to explore all the feasible state and action pairs to understand the network and attacks before the network state significantly changes or the attackers change their policies. It has been found that the learning speed of a Q-learning based scheme is usually slower than the network variation speed, which seriously degrades the edge security performance.

Therefore, the Dyna-Q based security methods such as the authentication scheme developed in \cite{Xiao2016PHY} use both the real defense experiences and the virtual experiences generated by the Dyna architecture to find the optimal strategy. {\color{black} The Dyna-Q based authentication scheme utilizes hypothetical experience to accelerate the learning process and thus improve the spoofing detection accuracy.}
However, the virtual experiences are not always true especially at the beginning of the security game, which decreases the learning rate of the security methods.

To address this issue, the edge security schemes based on post decision state (PDS) \cite{He2015Improving} apply the known information regarding the network, attack and channel models to accelerate the exploration and use Q-learning to study the unknown state space. On the other hand, the edge node without being aware of any network model can resort to the DQN technique that compresses the state space with deep learning. The DQN-based security schemes converge to the optimal strategies faster compared with the RL techniques mentioned above, especially when the edge node witnesses a large network state space. However, the implementation of the CNN in these schemes requires high computational complexity and memory, which exceeds the capability of many edge devices and mobile devices. {\color{black} To this end, a hotbooting method as a special case of transfer learning exploits the learning experiences in similar scenarios to initialize the weights of the CNN and reduce the random explorations at the beginning of the learning process. A fast DQN based anti-jamming communication method presented in \cite{Han2017Two} applies both DQN and the hotbooting technique to improve the communication efficiency against jamming.}
%To this end, we propose a hotbooting method as a special case of transfer learning that reuses the experiences in similar scenarios to initiate the CNN parameters and design a fast DQN based anti-jamming communication method in \cite{Han2017Two} to improve the communication efficiency against jamming.

Simulations and preliminary experiments built on laptops and USRPs show that the RL-based security solutions are promising to improve edge security. For example, we consider the mobile offloading of a user device with three edge candidates in the area against a mobile sweeping jammer. {\color{black} As presented in Fig. \ref{fig:ave_performance}, the DQN-based offloading scheme can significantly reduce the offloading energy consumption and the delay, and increase the SINR of the signals received by the edge nodes compared with the benchmark schemes.} All these schemes converge to the optimal strategy that can be validated via the Nash Equilibrium of the repeated edge security game after a sufficiently long time, although DQN requires the shortest learning time.

\section{Challenges \& Future Work}
Most existing RL techniques are firstly developed for various games, in which a learning agent accurately knows its state and immediate reward from each action (e.g., the change of the scores in a video game). In addition, an agent can tolerate most results of the feasible strategies especially at the beginning of the game, which is the basis of the trial-and-error methods. Unfortunately, these assumptions do not hold in network security. For instance, a non-optimal network defense decision sometimes leads to forbidding results such as national safety risks. Although the RL-based security techniques are promising to improve edge security and privacy, they have to address the following challenges.

\textbf{Inaccurate and delayed state information:} An edge device usually has difficulty estimating the current network and attack state accurately and fast enough to choose the next defense policy. Therefore, the impacts of the inaccurate and delayed state information on the MEC security performance have to be investigated. {\color{black} We have to improve the MEC security solutions with the advanced RL techniques that require less state information and tolerate the inaccurate and delayed state observation for 5G communication systems. A promising solution is to incorporate the known network and attack information extracted with data mining to accelerate the learning process.}

\textbf{Evaluation of the utility for each security strategy:} An agent has to observe the security gain and the protection cost to evaluate its reward from each action. Both in turn consists of a large number of factors. For example, in a secure mobile offloading, a mobile device has to accurately evaluate the data privacy, the transmission and computation delay, the energy cost and the rogue edge risks from its last offloading policy, and incorporates them properly to evaluate the utility, which is challenging for most practical MEC systems. {\color{black} The 5G communication systems have to investigate these factors in the utility evaluation instead of using the heuristic model used in most existing RL-based security schemes. It is critical to replace the heuristic RL methods such as Q-learning in the MEC security solutions with the newly developed RL techniques that work well with delayed and inaccurate utility information.}

\textbf{Makeup protocol for the bad RL decision:} Existing RL techniques require an agent to try some bad policies to learn the optimal strategy. This exploration that is dangerous for edge security indicates a large number of failed defense against attackers. To this end, transfer learning techniques that use data mining to explore existing defense experiences can be designed to help RL reduce the random exploration and thus decreases the risks of trying bad defense policies at the beginning of the learning process. {\color{black} Backup protocols have to be designed for the 5G system to avoid the security disaster from a bad decision made in the learning process such as connecting with a rogue edge.}

\section{Conclusions}
In this article, we studied several security challenges in MEC systems and proposed a security solution based on reinforcement learning. The solution consists of a secure mobile offloading solution against smart attacks, a light-weight authentication and a caching collaboration scheme to resist wiretaps. We applied RL to choose the defense levels and/or key parameters in the process. {\color{black} The RL-based secure mobile edge caching can enhance the security and user privacy of mobile edge caching systems.} As shown in the simulation results, the RL-based security solution is effective in protecting the MEC systems against various types of smart attacks with low overhead.

\ifCLASSOPTIONcaptionsoff
  \newpage
\fi

\bibliography{Caching}
\bibliographystyle{IEEEtr}

\balance
\end{document}